\documentclass[aps,prl,10pt,twocolumn,showpacs,preprintnumbers,amsmath,amssymb,showkeys,citeautoscript]{revtex4-1}
\usepackage{pdftexcmds}
\usepackage[version=3]{mhchem}
\usepackage{siunitx}
\usepackage{xcolor}
\usepackage{graphicx}
\usepackage{dcolumn}
\usepackage{bm}
\usepackage{amsmath}
\usepackage{tikz}
\usepackage{pgfplots}

\begin{document}
	
\title{Enabling visible-light absorption and $p$-type doping in \ce{In2O3} by adding Bi}
%\ce{(In_{1-$x$}Bi_{$x$})_2O3}}
	
%\author{Fernando P. Sabino}
%\affiliation{Department of Materials Science and Engineering, University of Delaware,
%Newark, Delaware 19716, USA}
%	
%\author{Su-Huai Wei}
%\affiliation{Beijing Computational Science Research Center, Beijing 100193, China}
%	
%\author{Anderson Janotti}
%\affiliation{Department of Materials Science and Engineering, University of Delaware,
%Newark, Delaware 19716, USA}

\author{Fernando P. Sabino$^1$, Su-Huai Wei$^2$, Anderson Janotti$^1$}
\affiliation{$^1$Department of Materials Science and Engineering, University of Delaware,
Newark, Delaware 19716, USA}
\affiliation{$^2$Beijing Computational Science Research Center, Beijing 100094, China}

\begin{abstract}

\ce{In2O3} is a prototype wide-band-gap semiconductor that exhibits metallic
conductivity when highly doped with \ce{Sn} while retaining a high degree of transparency
to visible light. It is widely used as a transparent window/electrode in solar
cells and LEDs. The functionality of \ce{In2O3} would be greatly extended if $p$-type conductivity
could also be achieved.
Using electronic structure calculations, we show that adding Bi to \ce{In2O3}, in the form
of dilute \ce{(In_{1-$x$}Bi_{$x$})_2O3} alloys, leads to a new valence band that is
sufficiently higher in energy than the original O-2$p$ valence band to allow for
$p$-type doping.
Moreover, the raised valence band in the \ce{(In_{1-$x$}Bi_{$x$})_2O3} dilute alloys leads to
strong optical absorption in the visible spectrum, opening up for new applications such as
a wide band gap absorber layer in tandem solar cells.

\end{abstract}


\maketitle

%%\section{Introduction} \label{sec:Introduction}

\ce{In2O3} is a wide-band-gap material, extensively used as a transparent conducting
oxide. When heavily doped with \ce{Sn}, often called ITO, it displays electrical conductivities
around \SI{1E3} {\siemens\per\centi\meter}, while preserving a high transparency
to visible light \cite{Walsh_167402_2008,Chen_13836_2013}. The combination of optical
transparency and high electrical conductivity is desirable for many
device applications such as liquid-crystal displays, OLEDs, touchscreens, and
transparent contacts in photovoltaic devices \cite{Nomura_1269_2003,Nomura_488_2004,Minami_S35_2005,Granqvist_1529_2007}.
The functionality of \ce{In2O3} would be greatly expanded if $p$-type
conductivity could also be achieved \cite{Xu_543_2000,Zhang_383002_2016}.

\begin{figure*}%[h][b]
\centering
\includegraphics[width=0.75\linewidth]{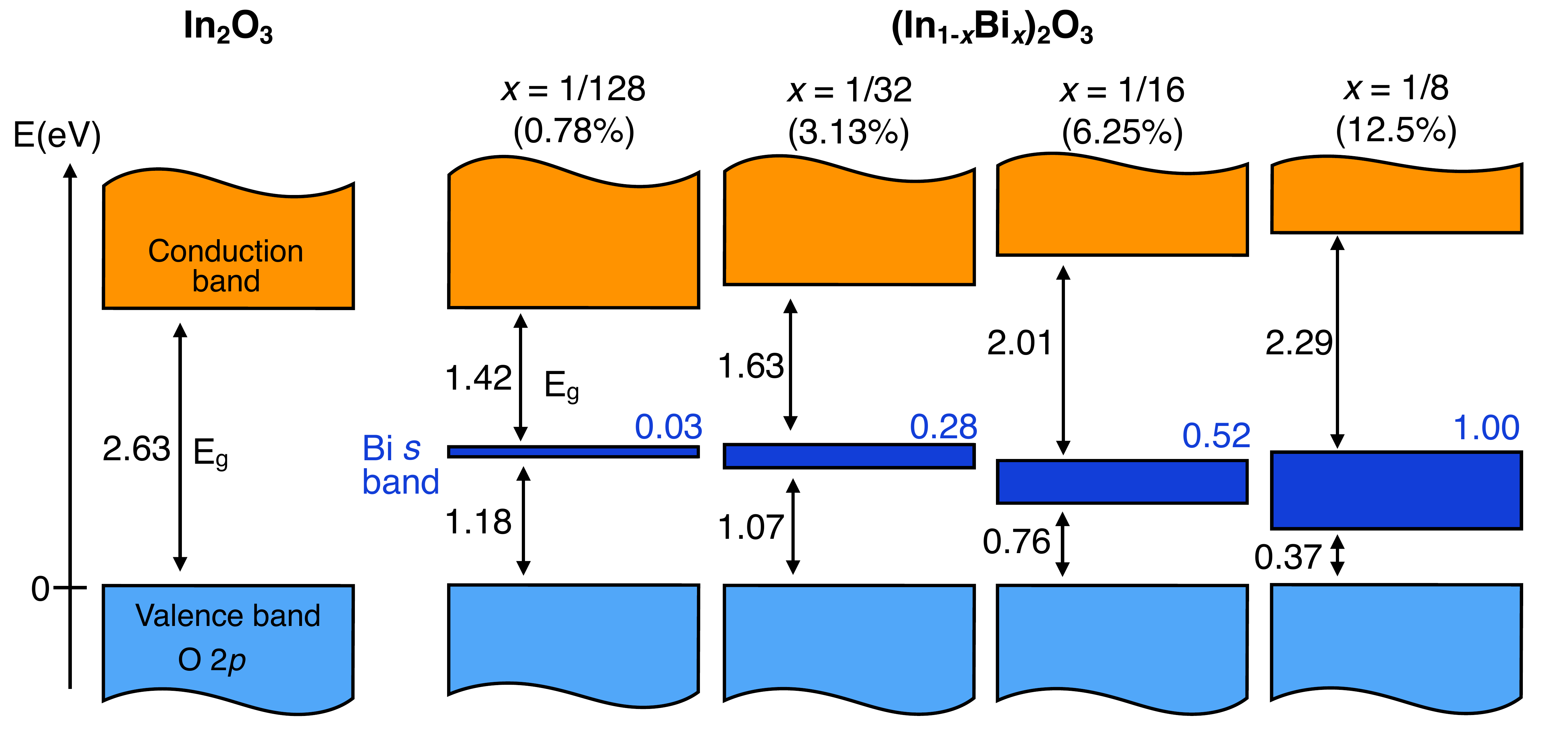}
\caption{
Position of the valence and conduction bands in \ce{In2O3} and dilute \ce{(In_{1-$x$}Bi_{$x$})_2O3} alloys for Bi concentrations of $x$=1/128 (0.78\%), 1/32 (3.13\%), 1/16 (6.25\%), and 1/8 (12.5\%).  We used the valence-band maximum (VBM) of \ce{In2O3} as reference and aligned the band edges of the alloys as described in the text.  The band width of the Bi $s$ band (valence band in the alloys) and the distance to the \ce{O} 2$p$ bands are indicated.
}
\label{band_alignment}
\end{figure*}

Most common dopants in \ce{In2O3} act as donors, including hydrogen \cite{Koida_L685_2007,Koida_033514_2010}.
This is explained by the relatively low position of the conduction band with respect
to the vacuum level. A major difficulty in making \ce{In2O3} $p$-type is related to
fundamental characteristics of its valence band \cite{Walsh_167402_2008,Sabino_205308_2015}.
It is composed of O 2$p$ orbitals and lies very low in energy with respect to the vacuum level, so that all possible acceptor
impurities lead to deep levels in the gap, with very high ionization energies \cite{Lany_085202_2009}.
Attempts to modify the electronic structure of wide-gap oxides to enable $p$-type
doping have been proposed, yet with limited success \cite{Kawazoe_939_1997,Zhang_383002_2016}.
The key strategy has been to modify the top of the valence band by enhancing covalence
through the orbital hybridization between O 2$p$ and metal cations \cite{Kawazoe_939_1997,Zhang_383002_2016},
or by incorporating isovalent anions, such as \ce{S} and \ce{Se}, with valence $p$ orbitals
that are higher in energy and more delocalized than the O 2$p$ \cite{Hiramatsu_125_2002,Hiramatsu_012104_2007}.
Building on recent ideas of using compounds containing post transition metals to modify the valence band \cite{Ogo_2187_2009,Yabuta_072111_2010,Bhatia_30_2016},
here we explore adding low concentrations of \ce{Bi} to \ce{In2O3}. We report on the stability,
as well as the electronic structure and optical properties of \ce{(In_{1-$x$}Bi_{$x$})_2O3}
dilute alloys.

Bismuth belongs to the column 5A of the periodic table. In III-V
semiconductors such as \ce{GaAs}, \ce{InAs}, and \ce{InSb}, \ce{Bi} substitute
on the pnictide sites, acting as an isovalent pentavalent addition.
However, \ce{Bi} is also known to behave as a trivalent species, such as in \ce{Bi2Se3} \cite{Shuk_179_1996}
and \ce{Bi2O3}  \cite{Walsh_235104_2006,Matsumoto_094117_2010}.
Considering the atomic size difference between In and O in \ce{In2O3}, we expect that Bi
will substitute on the In site.

Using density functional theory with a hybrid functional we calculate the electronic
structure of dilute \ce{(In_{1-$x$}Bi_{$x$})_2O3} alloys,  with $x$=\SI{0.78}{\percent}, 
\SI{3.13}{\percent}, \SI{6.25}{\percent} and \SI{12.5}{\percent}. We find that \ce{Bi} leads to an
occupied band, composed of \ce{Bi} $s$ and \ce{O} $p$ orbitals, that is significantly higher in energy than the valence-band maximum (VBM)
of the host material \ce{In2O3}. The band gap, as well as the position and bandwidth of
this new valence band can be tuned with \ce{Bi} concentration (see Figure~\ref{band_alignment}).
Furthermore, this raised valence band in \ce{(In_{1-$x$}Bi_{$x$})_2O3} dilute alloys not only leads to strong optical absorption in the visible spectral region, but also potentially enables $p$-type doping.

%\section{Theoretical Approach and Computational Retails}

Our first-principles calculations are based on the density functional theory \cite{Hohenberg_B864_1964,Kohn_A1133_1965} and the hybrid functional of Heyd-Scuseria-Ernzerhof (HSE06) \cite{Heyd_7274_2004,Heyd_219906_2006}, as implemented in the VASP code \cite{Kresse_13115_1993,Kresse_11169_1996}.
For the interaction between the valence electrons
and the ionic cores, we used augmented plane wave (PAW) potentials \cite{Blochl_17953_1994,Kresse_1758_1999},
with the following valence configurations: \ce{O}($2s^{2}2p^{4}$), \ce{In}($4d^{10}5s^{2}5p^{1}$)
and \ce{Bi}($5d^{10}6s^{2}6p^{3}$).
The volume and atomic positions were relaxed using the the semi-local exchange and correlation functional
of Perdew, Burke, and Ernzerhof parameterized for solids (PBESol) \cite{Perdew_136406_2008},
with an energy cutoff of \SI{620}{\electronvolt} for the plane-wave expansion. For integrations over the Brillouin zone, we employed a
$3{\times}3{\times}3$ \textbf{k}-point mesh for the 40-atom primitive cell of bixbyite \ce{In2O3}, and the same \textbf{k}-points density for the other supercells.

Since PBE or PBESol drastically underestimate band gaps \cite{Perdew_497_1985,Heyd_7274_2004,Heyd_219906_2006}, we use the HSE06 to calculate the electronic structure of \ce{In2O3} and \ce{(In_{1-$x$}Bi_{$x$})_2O3}
alloys.  The HSE06 calculations  were carried out for the crystal structures optimized with PBESol.
In HSE06, the nonlocal Fock exchange is separated in short
and long-range parts, by a screening function determined by the parameter $\omega = 0.206$ \cite{Heyd_7274_2004,Heyd_219906_2006}.
In the short-range region we used 25\% of exact exchange, and 75\% of PBE exchange, while the long range part
is described by the PBE exchange \cite{Heyd_7274_2004,Heyd_219906_2006}.
The electronic band structure and density of states were computed
with a lower cutoff energy of \SI{470}{\electronvolt} and
a large $7{\times}7{\times}7$ \textbf{k}-point mesh. For the optical properties
we use the tetrahedral smearing method, but do not include excitonic effects due to the prohibitive computational cost.

%\section{Results}

%\subsection{Crystal structures and relatively stability}

\ce{In2O3} crystallizes in the body centered cubic bixbyite structure \cite{Marezio_723_1966} with
space group $Ia\overline{3}$ and having 40 atoms (8 formula units) per primitive cell.
All the \ce{O} sites are equivalent, and each \ce{O} atom is bonded to four \ce{In}
atoms, whereas each \ce{In} atom is bonded to six \ce{O} atoms. There are two
inequivalent \ce{In} octahedral sites, one forming a distorted octahedron, and the other forming a
perfect octahedron (all \ce{In}-\ce{O} bond lengths are equal). Therefore, there
are two possible sites for \ce{Bi} to substitute for \ce{In} in the bixbyite
structure. We tested \ce{Bi} substitution on both sites, and we find that \ce{Bi}
occupying the perfect octahedral site is only \SI{12}{\milli\electronvolt} lower in energy
than the distorted site for \ce{(In_{1-$x$}Bi_{$x$})_2O3} with $x=1/16$ (6.25\%).
The difference in the electronic structure of these two configurations is negligible, with band gaps differing
by less than  \SI{10}{\milli\electronvolt}
Therefore, we expect that \ce{Bi} will incorporate on both perfect and distorted octahedral sites.

\ce{(In_{1-$x$}Bi_{$x$})_2O3} structures were built using one Bi in supercells of various sizes, from 40-atom to
320-atom, simulating dilute \ce{Bi} concentrations of 0.78\%, 3.13\%, 6.25\%, and 12.5\%.
% were build based on $1{\times}1{\times}1$
%and $2{\times}2{\times}2$ primitive, and $1{\times}1{\times}1$ conventional
%cubic bixbyite.
We also tested using special quasi random structures (SQS) \cite{Zunger_353_1990,Wei_9622_1990}to simulate the alloys.
The results were quite similar, with band gaps differing by less that 0.1 eV.

\begin{table}[t!]
\centering
\caption{Calculated lattice parameter $a_0$, volume, and mixing enthalpy ($\Delta H^f$) per cation
of  \ce{(In_{1-$x$}Bi_{$x$})_2O3} alloys for various \ce{Bi} concentrations ($x$).
For comparison, the experimental equilibrium lattice parameter of  \ce{In2O3} and volume per formula unit are 10.12~\AA, and 64.78~\AA$^3$ \cite{Marezio_723_1966}.
}
\label{lattice}
\begin{ruledtabular}
\begin{tabular}{lcccc}
System                        &~$x$              & $\Delta H^f$            & $a_0$          & Volume         \\
                              &($\%$)            & (meV/cation)              & ($\rm{\AA}$)   & ($\rm{\AA}^3$/f.u.)  \\
\hline
\ce{In2O3}                    & ~~~$0$          & ~~$0$                           & 10.14          & 65.15        \\
\ce{(In_{1-$x$}Bi_{$x$})_2O3} & ~$0.78$          & ~~2                  & 10.15          & 65.33         \\
\ce{(In_{1-$x$}Bi_{$x$})_2O3} & ~$3.13$          & ~13                  & 10.17          & 65.74         \\
\ce{(In_{1-$x$}Bi_{$x$})_2O3} & ~$6.25$          & ~25                  & 10.20          & 66.33          \\
\ce{(In_{1-$x$}Bi_{$x$})_2O3} & ~$12.5$           & ~52                  & 10.26          & 67.49      \\
%\ce{In2O3} (exp.)                   & -    & -                                              & 10.12           & 64.78   \\
\end{tabular}
\end{ruledtabular}
\end{table}

The calculated lattice parameter $a_0$ and volume per formula unit of \ce{In2O3} and \ce{(In_{1-$x$}Bi_{$x$})_2O3} alloys are listed in Table~\ref{lattice}. The calculated lattice parameter of 10.14~\AA for \ce{In2O3}, using PBEsol, is only 0.2\% larger than the experimental value of 10.12~\AA \cite{Marezio_723_1966}.
Adding \ce{Bi} to \ce{In2O3} leads to an increase in the lattice parameter since
\ce{Bi} has a larger atomic radius than \ce{In}.
We also calculated the mixing enthalpy for the \ce{(In_{1-$x$}Bi_{$x$})_2O3} alloys,
which we define as:
\begin{eqnarray}
\Delta H^f & = E_{tot}[({\rm In}_{1-x}{\rm Bi}_{x})_2{\rm O}_3] -(1-x)E_{tot}({\rm In}_2{\rm O}_3) \nonumber \\
            &  -xE_{tot}({\rm Bi}_2{\rm O}_3),
\end{eqnarray}
where $E_{tot}[({\rm In}_{1-x}{\rm Bi}_{x})_2{\rm O}_3]$ is the total energy of the alloy,
$E_{tot}({\rm In}_2{\rm O}_3)$ is the total energy of \ce{In2O3} and $E_{tot}({\rm Bi}_2{\rm O}_3)$
is the total energy of \ce{Bi2O3} in its ground state.
The calculated formation enthalpy of the \ce{(In_{1-$x$}Bi_{$x$})_2O3} alloys are comparable to that
found in \ce{(Ga_{$1-x$}In_{$x$})_2O3} \cite{Peelaers_085206_2015}, which have been demonstrated experimentally \cite{Baldini_552_2014}.
For instance, $\Delta H^f $= 52 meV/cation for \ce{(In_{1-$x$}Bi_{$x$})_2O3} with $x$=12.5\%, whereas  $\Delta H^f $= 50 meV/cation for
\ce{(Ga_{$1-x$}In_{$x$})_2O3} for the same $x$ \cite{Peelaers_085206_2015}.
Therefore, these relatively low values of mixing enthalpies indicate that \ce{(In_{1-$x$}Bi_{$x$})_2O3} alloys are likely to form.

\begin{figure}
\centering
\includegraphics[width=1.0\linewidth]{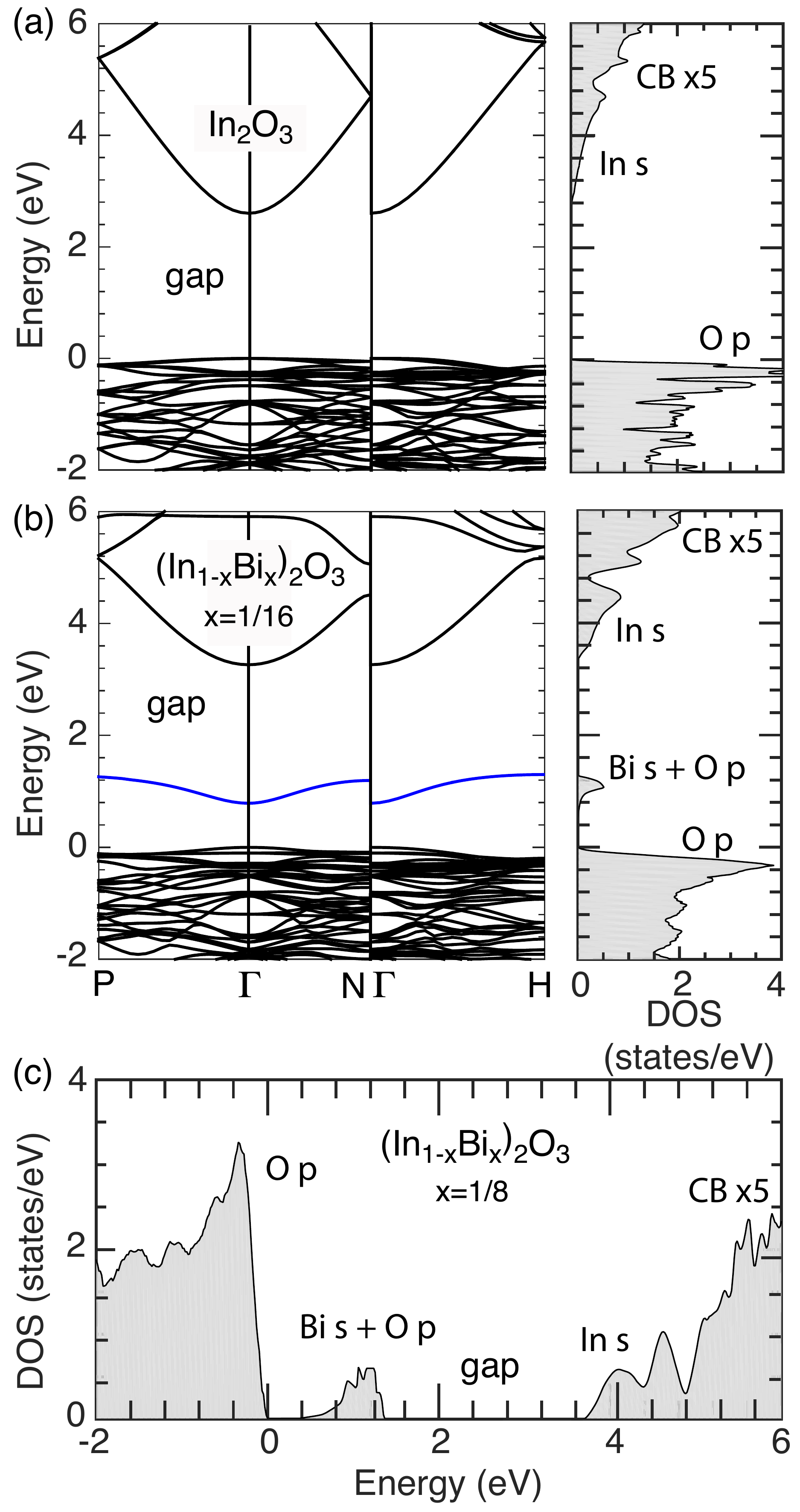}
\caption{Calculated band structure and density of states for (a)  \ce{In2O3}  and (b) \ce{(In_{1-$x$}Bi_{$x$})_2O3} with $x$=1/16, and (c) \ce{(In_{1-$x$}Bi_{$x$})_2O3} with $x$=1/8 aligned using the averaged electrostatic potential over a In atom in the bulk \ce{In2O3} and in a In atom far from the Bi in the \ce{(In_{1-$x$}Bi_{$x$})_2O3} alloy.  The valence-band maximum (VBM) of  \ce{In2O3} was used as reference, and indicates that the VBM of the alloy is 1.28 eV ($x$=1/16) and 1.37 eV ($x$=1/8) higher than the original VBM of the host material.
}
\label{band}
\end{figure}

The calculated electronic band structure and density of states (DOS) of \ce{In2O3} and \ce{(In_{1-$x$}Bi_{$x$})_2O3} alloy with $x=1/16$ and $x=1/8$ are shown in Figure~\ref{band}. Note that, for $x=1/16$ an "ordered" structure was used to model the \ce{(In_{1-$x$}Bi_{$x$})_2O3} alloy which maximize the Bi-Bi distance. In practice, we expect that random distribution of \ce{Bi} would lead to a broadening of the band. Thus for $x=1/8$, which \ce{Bi} occupation was determined by SQS, we plot only the DOS.

%Note that in practice, we expect that random distribution of Bi would lead to a lack of long range order, and therefore, a Brillouin zone that would collapse to the $\Gamma$ point. However, it still
%instructive to analyze the band structure of the (ordered) \ce{(In_{1-$x$}Bi_{$x$})_2O3} alloy.

The band structure and the DOS
for \ce{In2O3}, calculated using HSE06, shows a
fundamental band gap of \SI{2.63}{\electronvolt}, which is slightly lower than the previously reported value of \SI{2.80}{\electronvolt} \cite{Walsh_167402_2008}.
Adding Bi leads to a fully occupied band which is significantly higher in energy than
the original VBM of the host \ce{In2O3} material. This band originates from the coupling between
the Bi $s$ state, which is well below the VBM, with O 2$p$ states, mainly from the O atoms next to the Bi.
Therefore, it has an antibonding character from the Bi $s$ O $p$ coupling.  The minimum occurs at $\Gamma$
and the maximum (VBM) occurs at the H point for the ordered alloy at $x=1/16$. Note that, in a random alloy as it is modeled 
for \ce{(In_{1-$x$}Bi_{$x$})_2O3} with $x=1/8$, the plot of the density of states as shown in Fig.~\ref{band}(c) is more relevant. 

For the \ce{(In_{1-$x$}Bi_{$x$})_2O3} alloy with $x=1/16$, we find a band gap of 2.01 eV, with a new VBM
about 1.28 eV higher than the VBM of \ce{In2O3}. The band width of this higher valence band is 0.52 eV.
The conduction-band minimum of the alloy, at $\Gamma$, is also pushed up by 0.66 eV, largely due to the interaction with the Bi $s$ band.
When the \ce{Bi} concentration increase to $x=1/8$, the band gap increased to 2.29 eV, with the antibonding \ce{Bi} $s$ band width of
1.00 eV and the CBM is pushed up by 1.03 eV.

The maximum of the O 2$p$ band in the \ce{(In_{1-$x$}Bi_{$x$})_2O3} alloy occurs at approximately the same energy
position as the VBM of the host \ce{In2O3}, which is composed of \ce{O} 2$p$.
The evolution of the band structure of the dilute \ce{(In_{1-$x$}Bi_{$x$})_2O3} alloy with Bi concentration,
shown in Figure~\ref{band_alignment}, indicates that we can tune three key parameters in the \ce{(In_{1-$x$}Bi_{$x$})_2O3} alloy
by changing the Bi concentration: the positions of the valence-band (now derived from the antibonding Bi $s$ state) and conduction-band edges,
the band width of the valence band, and the band gap.

The repulsion between the \ce{Bi} $s$ valence band and the lowest energy conduction
band also plays an important role in determining the band gap of the alloy.
While the fundamental gap is direct at $\Gamma$ in \ce{In2O3}, it becomes
indirect in the (ordered) alloy, with the VBM at H-point for $x = 1/16$ and $1/128$,
and at the R-point for $x = 1/32$. For the random alloys $x=1/8$, the band gap has a pseudodirect nature.

%\subsection{Optical Properties}

\begin{figure}
\centering
\includegraphics[width=1.0\linewidth]{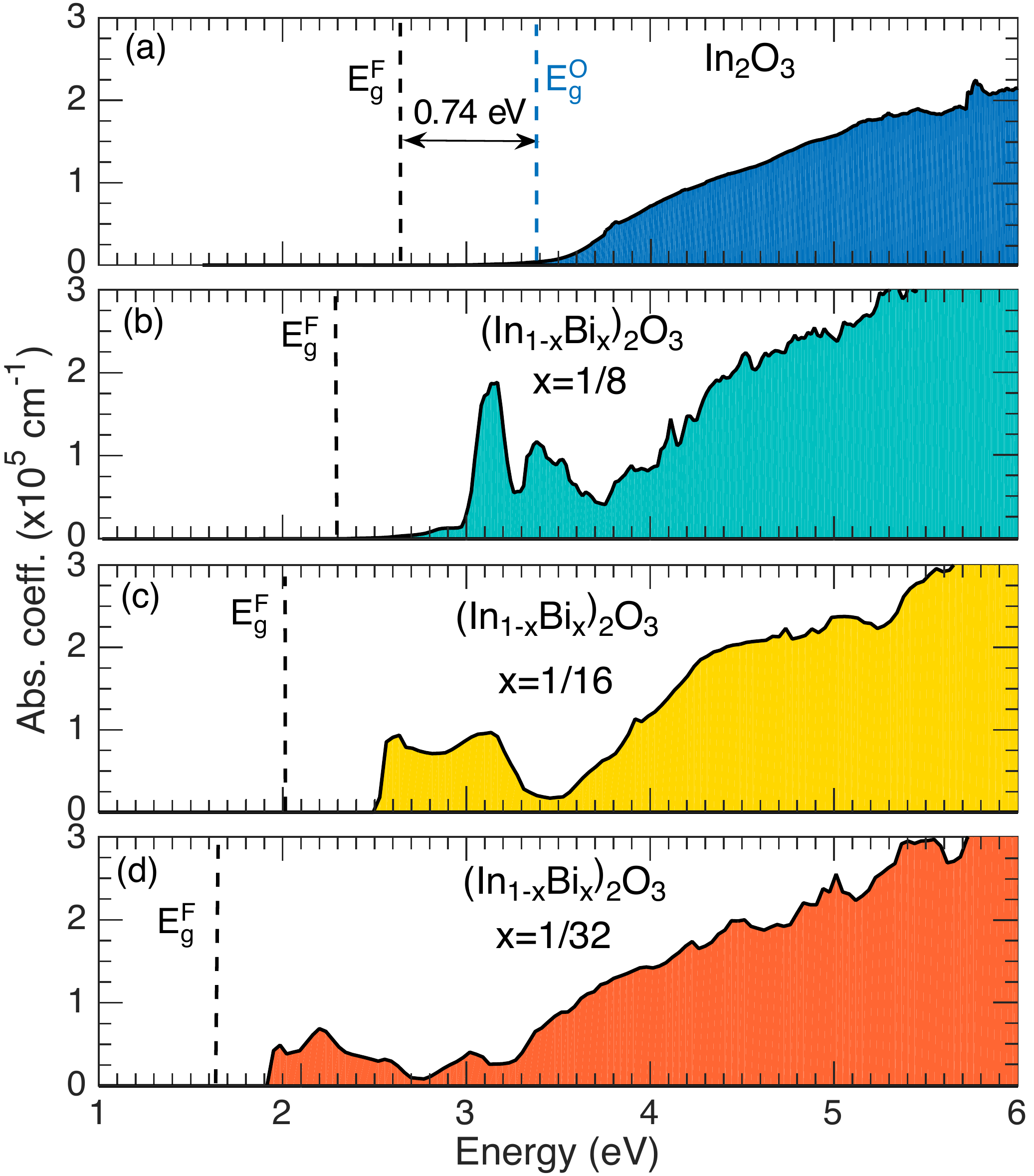}
\caption{
Calculated absorption coefficient as a function of photon energy for \ce{In2O3} and
\ce{(In_{1-$x$}Bi_{$x$})_2O3} with Bi concentrations $x = 1/8$, $x = 1/16$ and $x = 1/32$.
The vertical dashed lines indicate the fundamental band gap ($E_g^F$) and the optical band gap ($E_g^O$). The disparity between $E_g^F$ and $E_g^O$ is also indicated.
}
\label{optical}
\end{figure}

We also calculated the absorption coefficient of dilute \ce{(In_{1-$x$}Bi_{$x$})_2O3} alloys.
Due to the high computational cost of calculating the optical properties
of large systems with sufficient number of \textbf{k}-points in HSE06, we performed
these calculations using PBESol and then applied a scissors operator to correct
the fundamental band gap according to the HSE06 results. This approximation does
not affect the lowest energy optical transitions, as tested
using HSE06 and PBESol for \ce{(In_{1-$x$}Bi_{$x$})_2O3} with $x = 1/16$.

The optical properties of \ce{In2O3} are well described in the literature,
including the disparity between the optical and fundamental band gap \cite{Walsh_167402_2008,Sabino_205308_2015}.
As shown in Figure~\ref{optical}(a), our results reproduce
previous calculations for \ce{In2O3}, where the optical absorption starts to increase
only at photon energies that are \SI{~0.8}{\electronvolt} larger than the fundamental
band gap $E_g^{F}$, defining the optical band gap $E_g^{O}$. Optical transitions
with energy in the range from $E_g^{F}$ to $E_g^{O}$ have negligible contributions to
the optical absorption coefficient, with magnitude smaller than \SI{1e-4} as
indicated in Figure~\ref{optical}. This small contribution comes from the weak
transitions from valence-band to conduction-band states at or in the vicinity of the
$\Gamma$ point \cite{Sabino_085501_2017}.

As shown in Figure~\ref{optical}(b)-(d), we see a red shift in the onset
of absorption of  \ce{(In_{1-$x$}Bi_{$x$})_2O3} alloys. This red shift
decreases as the Bi concentration increases, consistent with the variation
of the band gap with Bi concentration shown in Figure~\ref{band_alignment}
and Figure~\ref{band}.  The peak associated with the transition from the antibonding
\ce{Bi} $s$ valence band to the conduction band increases with the \ce{Bi} concentration.
We note, however, a seeming disparity between the optical and fundamental band gap
in the \ce{(In_{1-$x$}Bi_{$x$})_2O3} alloys, as shown in Figure~\ref{optical}(b)-(d).
The origin of this disparity is associated to the indirect band gap in the alloys simulated
using periodic supercells.  In a real random alloy,
the lack of long range order would lead to the collapse of the Brillouin zone to the $\Gamma$ point,
and the discussion of direct versus indirect band gap will lack meaning, so that we would expect
transitions near or at the fundamental band gap to be allowed, but perhaps weak.

%There is also a strong enhancement in the optical
%absorption near the onset, due to the transitions from the \ce{Bi} $s$ band to the conduction
%band when compared to the optical absorption in pure \ce{In2O3}.

We note that the VBM in the \ce{(In_{1-$x$}Bi_{$x$})_2O3} alloys is sufficiently higher in energy
than the VBM in  \ce{In2O3} to allow for $p$-type doping.
 For example, introducing Mg dopants in \ce{In2O3} leads
to a deep acceptor level in the gap associated with the hole localized on an O atom next to the Mg$_{\rm In}$
substitutional impurity \cite{Lany_085202_2009}. This acceptor level is about 0.9 eV above the VBM in \ce{In2O3}.
Therefore, in dilute \ce{(In_{1-$x$}Bi_{$x$})_2O3} alloys, the same acceptor level is expected
to appear close to the antibonding Bi-$s$-derived VBM in the alloy (see band alignment in Figure~\ref{band_alignment}),
making it a shallow acceptor.

Finally, we believe our results can explain some puzzling results on \ce{In2O3} samples containing Bi.
A high intensity emission peak at \SI{1.8}{\electronvolt} in the spectrum of
nanocomposites of \ce{Bi}-doped ITO has been reported by Na {\em et al.} \cite{Na_095016_2016}.
The authors have attributed this emission to a transition from the conduction band
to deep levels related to \ce{O} vacancies. However, based on our results, for
\ce{Bi} concentration of approximately 1\% in their experiments, the predicted optical band gap of the dilute
\ce{(In_{1-$x$}Bi_{$x$})_2O3} alloy is \SI{~1.8}{\electronvolt}, and a transition from conduction band to the antibonding
\ce{Bi} $s$ band can explain this emission peak.

Considering the strong optical absorption in the visible spectrum of dilute \ce{(In_{1-$x$}Bi_{$x$})_2O3}
alloy, the conduction-band offset with \ce{In2O3}, and the possibility of doping it $p$-type, we can think of
\ce{(In_{1-$x$}Bi_{$x$})_2O3} as a wide band gap absorption layer to be integrated in tandem solar cells where
ITO is used as transparent contact. Note that there will be a tradeoff between band gap and the width of the antibonding
Bi $s$ valence band, which will have to be taken into consideration in the device design.

%\section{Conclusion}

In summary, using electronic structure calculations, we show that incorporating low
concentrations of \ce{Bi} into \ce{In2O3}, forming dilute \ce{(In_{1-$x$}Bi_{$x$})_2O3}
alloys, leads to significant changes in the electronic and optical properties
of the parent compound, greatly extending its functionality. The \ce{Bi} introduces
an occupied intermediate valence band, composed mainly of \ce{Bi} $s$ and \ce{O} $p$, resulting
in significantly reduced band gaps. The position and band width of the new valence
band in the \ce{(In_{1-$x$}Bi_{$x$})_2O3} alloys can be tuned by controlling the
\ce{Bi} concentration. The high VBM of the \ce{(In_{1-$x$}Bi_{$x$})_2O3} alloys can possibly enable $p$-type doping. The reduced band gap, compared to
\ce{In2O3} and strong near band edge absorption make dilute \ce{(In_{1-$x$}Bi_{$x$})_2O3}
 alloys a promising wide-gap absorber layer in tandem solar cells.

\section*{Acknowledgments}

This work at University of Delaware was supported by the National Science Foundation Faculty Early Career Development Program DMR-1652994. The work of SHW at CSRC was supported by the National Nature Science Foundation of China under Grant No. 11634003 and U1530401. This research was also supported by the the Extreme Science and Engineering Discovery Environment supercomputer facility, National Science Foundation grant number ACI-1053575, and the Information Technologies (IT) resources at the University of Delaware, specifically the high performance computing resources.


%\bibliography{jshort_25Feb2015,boxref_25Feb2015}

\end{document}